\documentclass[12pt]{article}
\usepackage{graphicx}

\begin{document}
\begin{center}

{\large\bf Excitation functions of parameters extracted from
three-source (net-)proton rapidity distributions in Au-Au and
Pb-Pb collisions over an energy range from AGS to RHIC}

\vskip1.0cm

Li-Na Gao$^{a}$, Fu-Hu Liu$^{a,}${\footnote{E-mail:
fuhuliu@163.com; fuhuliu@sxu.edu.cn}}, Yan Sun$^{b}$, Zhu
Sun$^{b}$, Roy A. Lacey$^{c}$,

{\small\it $^a$Institute of Theoretical Physics, Shanxi
University, Taiyuan, Shanxi 030006, China

$^b$Department of Physics, Shanxi Datong University, Datong,
Shanxi 037009, China

$^c$Departments of Chemistry \& Physics, Stony Brook University,
Stony Brook, NY 11794, USA}
\end{center}

\vskip1.0cm

{\bf Abstract:} Experimental results of the rapidity spectra of
protons and net-protons (protons minus antiprotons) emitted in
gold-gold (Au-Au) and lead-lead (Pb-Pb) collisions, measured by a
few collaborations at the alternating gradient synchrotron (AGS),
super proton synchrotron (SPS), and relativistic heavy ion
collider (RHIC), are described by a three-source distribution. The
values of the distribution width $\sigma_C$ and fraction $k_C$ of
the central rapidity region, and the distribution width $\sigma_F$
and rapidity shift $\Delta y$ of the forward/backward rapidity
regions, are then obtained. The excitation function of $\sigma_C$
increases generally with increase of the center-of-mass energy per
nucleon pair $\sqrt{s_{NN}}$. The excitation function of
$\sigma_F$ shows a saturation at $\sqrt{s_{NN}}=8.8$ GeV. The
excitation function of $k_C$ shows a minimum at
$\sqrt{s_{NN}}=8.8$ GeV and a saturation at $\sqrt{s_{NN}}\approx
17$ GeV. The excitation function of $\Delta y$ increases linearly
with $\ln(\sqrt{s_{NN}})$ in the considered energy range.
\\

\\


\vskip1.0cm

{\section{Introduction}}

Rapidity (or pseudorapidity) distributions of charged particles
produced in nucleus-nucleus (heavy ion) collisions at high
energies are an important topic measured in experiments and
analyzed in theories [1--3]. In particular, the rapidity
distributions of protons and net-protons (protons minus
antiprotons, i.e. $p-\bar p$) are related to not only the energy
loss or stopping power of colliding nuclei but also the number
density and mean free path of the considered particles, which are
related to the onset of deconfinement in high energy
nucleus-nucleus collisions [4--6]. This energy loss of colliding
nuclei is also a fundamental quantity which determines the
effective energy available for production yields of particles and
excitation degree of interacting system [7]. This effective energy
is essential for the formation of a deconfined quark and gluon
phase of matter, i.e. quark-gluon plasma (QGP) or quark matter. To
determine the critical energy of phase transition from hadronic
matter to QGP is an important and difficult topic [8]. Before
making a determination, one needs to search for some softest
points in the equation of state (EoS), which should be relative to
the local minimums in some excitation functions which are the
dependences of parameters on collision energy.

The width and peak (or hollow) position of rapidity distribution
are very sensitive quantities which are related to collision
energy, in particular for peak (or hollow) position in the
laboratory reference frame. It is expected that the system at
different collision energies corresponds to different interacting
mechanisms. In nucleus-nucleus collisions over an energy range
from the alternating gradient synchrotron (AGS) to super proton
synchrotron (SPS) and relativistic heavy ion collider (RHIC), the
interacting system undergoes the liquid-like state of nucleon
matter, gas-like state of hadronic matter, and liquid-like sate of
quark matter. From AGS to RHIC, it is expected that the excitation
functions of the mentioned parameters should experience a minimum
or other change due to the interacting system changing from
liquid-like state to gas-like state and then liquid-like state
again, though the two liquid-like states correspond to different
types of matters.

Although we can simply and approximately use a Gaussian function
to fit rapidity distributions in some cases, two or more Gaussian
functions can give better descriptions. In fact, in most cases,
the experimental rapidity spectra cannot be exactly described by a
single Gaussian function [9, 10]. Instead, we need two or three
Gaussian functions to describe the experimental spectra. In the
case of using a two-Gaussian function [9, 10], the first and
second functions correspond to the contributions of the forward
and backward (or backward and forward) rapidity regions
respectively. In the case of using a three-Gaussian function, the
third function is regarded as the contribution of the central
rapidity region [4--6, 11--13]. Since the two-source is possibly
misunderstood by the false appearance which implies an
unbelievable case that there is no source in the central rapidity
region, we would rather use the three-source distribution [4--6,
11--13] than the two-source one, though the former has more
parameters than the latter.

Although the three-source distribution has been used in our
previous works [14--16], in which the rapidity distributions of
charged particles, which are mainly charged pions, produced in
proton-proton and nucleus-nucleus collisions were reported, the
present work focusses on the rapidity spectra of (net-)protons
emitted in gold-gold (Au-Au) and lead-lead (Pb-Pb) collisions over
an energy range from AGS to RHIC. It is expected that some
minimums and other phenomenons in the excitation functions of
parameters extracted from the rapidity spectra of (net-)protons
should be observed.

The rest of this paper is structured as follows. The model and
formalism are shortly described in section 2. Results and
discussion are given in section 3. In section 4, we summarize our
main observations and conclusions.
\\

{\section{The model and formalism}}

As introduced above, we would like to use three Gaussian functions
[14--16] to fit the rapidity spectra of charged particles produced
in nucleus-nucleus collisions at high energies, where the charged
particles are mainly charged pions, and include also protons and
others. The three Gaussian functions cater to the three-source
model [4--6, 11--13]. Although the three-source distribution uses
more parameters than the two-source one, the former should give a
better description than the latter.

Based on the focus of the present work and the three-source
distribution [4--6, 14--16], we assume the rapidity distribution,
$dN/dy$, of (net-)protons emitted in symmetric nucleus-nucleus
collisions at high energies to be
\begin{eqnarray}
\frac{dN}{dy} &=& \frac{N_0}{\sqrt{2\pi}} \Bigg\{
\frac{1-k_C}{2\sigma_F} \exp \bigg[- \frac{(y-y_{cm}+\Delta
y)^2}{2\sigma_F^2} \bigg]
\nonumber\\[0.1cm]
&&+ \frac{k_C}{\sigma_C} \exp \bigg[-
\frac{(y-y_{cm})^2}{2\sigma_C^2} \bigg]
\nonumber\\[0.1cm]
&&+ \frac{1-k_C}{2\sigma_F} \exp \bigg[- \frac{(y-y_{cm}-\Delta
y)^2}{2\sigma_F^2} \bigg] \Bigg\},
\end{eqnarray}
where $y_{cm}$ denotes the mid-rapidity or the rapidity of the
center-of-mass, and $y_{cm}=0$ for symmetric collisions in the
center-of-mass reference frame; $\sigma_C$ and $k_C$ denote the
rapidity distribution width and fraction of the central rapidity
region, respectively; $\sigma_F$ and $\Delta y$ denote the
rapidity distribution width and rapidity shift (peak position) of
the forward/backward rapidity regions, respectively; and $N_0$
denotes the normalization constant when we compare the normalized
distribution to the experimental spectrum. In the present work,
$y$ and $y-y_{cm}$ are in fact the rapidities in the laboratory
and center-of-mass reference frames, respectively.

We have four main parameters, $\sigma_C$, $\sigma_F$, $k_C$, and
$\Delta y$, in the above discussion. The selections of the four
main parameters are influenced each other in the fitting process.
To determine the best values of parameters, we have used the
method of least squares. The minimum $\chi^2$ corresponds to the
best values of parameters. An appropriate increase in $\chi^2$
determines the uncertainties of parameters when we increase or
decrease some amounts in the best values of parameters. By
analyzing experimental data in nucleus-nucleus collisions over an
energy range from AGS to RHIC as much as possible, we can obtain
the excitation functions of the four parameters in the considered
energy range. At the same time, the dependences of parameters on
collision centrality can be incidentally studied.
\\

\begin{figure}
\hskip-1.0cm \begin{center}
\includegraphics[width=15.0cm]{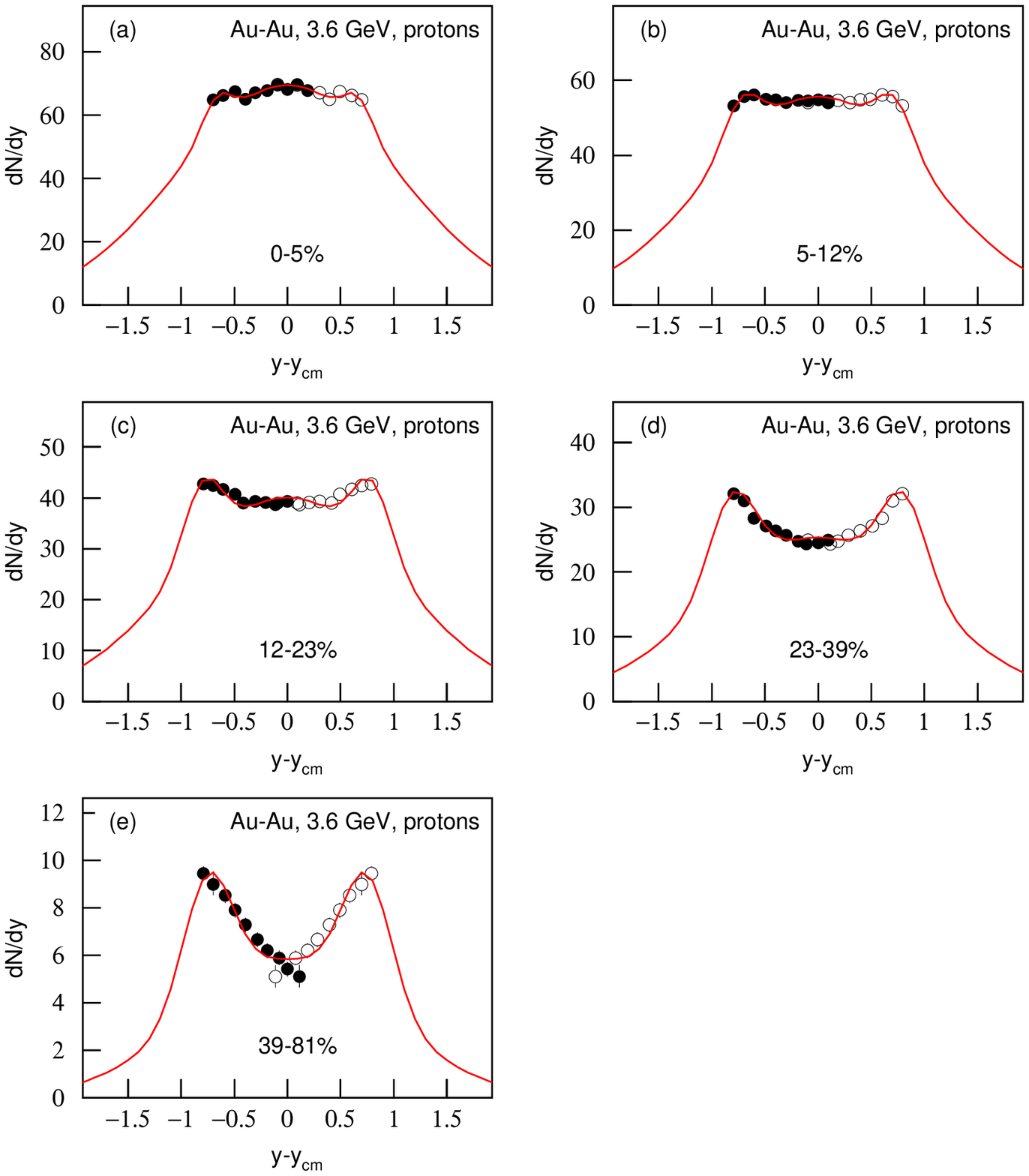}
\end{center}
\vskip.0cm\small Figure 1. Rapidity distributions of protons
emitted in Au-Au collisions at $\sqrt{s_{NN}}=3.6$ GeV, where
panels (a)--(e) correspond to centrality intervals 0--5\%,
5--12\%, 12--23\%, 23--39\%, and 39--81\%, respectively. The
closed circles represent the experimental data of the E917
Collaboration [17] and the open circles are reflected at the
mid-rapidity. The curves are our results fitted by the
three-source distribution.
\end{figure}

\begin{figure}
\hskip-1.0cm \begin{center}
\includegraphics[width=15.0cm]{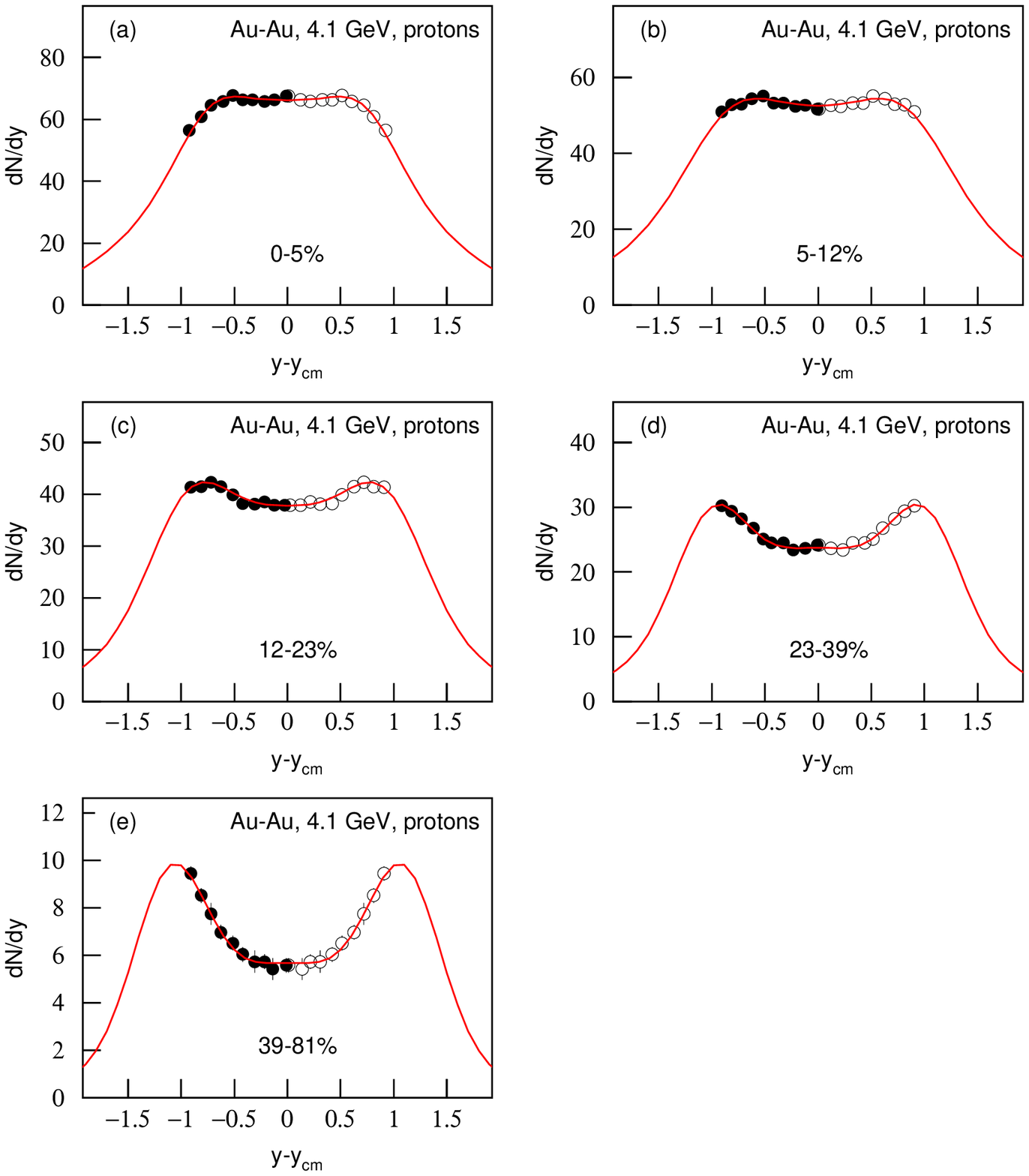}
\end{center}
\vskip.0cm\small Figure 2. The same as Figure 1, but showing the
results at $\sqrt{s_{NN}}=4.1$ GeV.
\end{figure}

\begin{figure}
\hskip-1.0cm \begin{center}
\includegraphics[width=15.0cm]{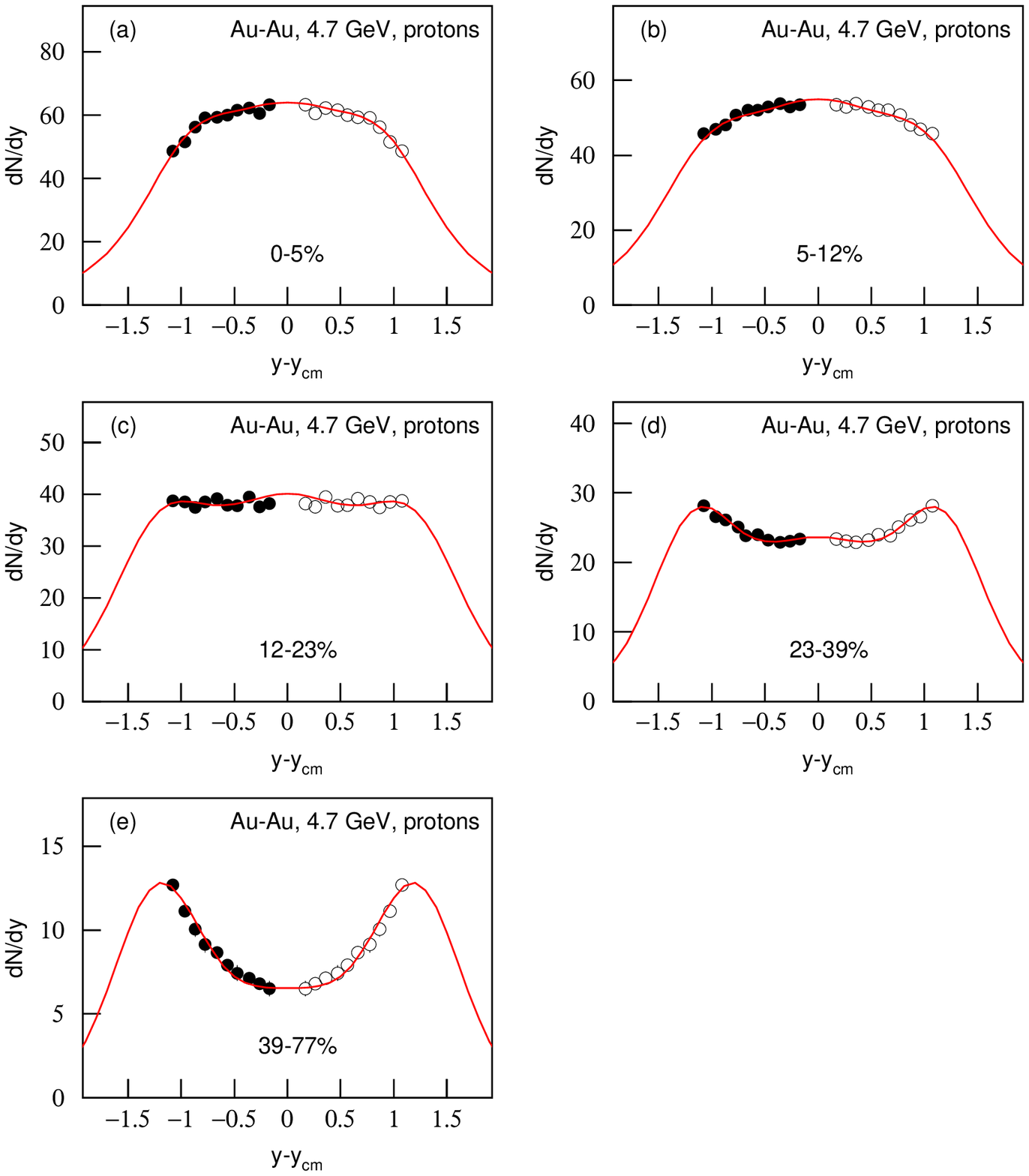}
\end{center}
\vskip.0cm\small Figure 3. The same as Figure 1, but showing the
results at $\sqrt{s_{NN}}=4.7$ GeV.
\end{figure}

{\section{Results and discussion}}

Figures 1--3 present the rapidity distributions of protons emitted
in Au-Au collisions at $\sqrt{s_{NN}}=3.6$, 4.1, and 4.7 GeV,
respectively. From panels (a) to (e), the corresponding centrality
intervals are 0--5\%, 5--12\%, 12--23\%, 23--39\%, and 39--81\%,
respectively. The closed circles represent the experimental data
of the E917 Collaboration [17] and the open circles are reflected
at the mid-rapidity (with $y-y_{cm}=0$). The curves are our
results fitted by the three-source distribution. The values of
free parameters $\sigma_C$, $\sigma_F$, $k_C$, and $\Delta y$, as
well as the normalization constant $N_0$ are listed in Table 1
with the values of $\chi^2$ per degree of freedom (dof). One can
see that the model describes approximately the considered
experimental data in most cases.

\vskip2.0cm

{\tiny {Table 1. Values of free parameters, normalization
constants, and $\chi^2$/dof corresponding to the curves in Figures
1--7, where Figures 5(b), 6, and 7 correspond to Pb-Pb collisions,
and others correspond to Au-Au collisions.
{%
\begin{center}
\begin{tabular}{cccccccccc}
\hline Figure & $\sqrt{s_{NN}}$ (GeV) & Centrality (\%) & Particle & $\sigma_C$ & $\sigma_F$ & $k_C$ & $\Delta y$ & $N_0$& $\chi^2$/dof \\
\hline
1(a) & 3.6  & 0--5   & $p$ & $1.030\pm0.017$ & $0.133\pm0.007$ & $0.965\pm0.010$ & $0.674\pm0.036$ & $186.2\pm0.8$ & 0.579 \\
1(b) & 3.6  & 5--12  & $p$ & $1.032\pm0.015$ & $0.166\pm0.008$ & $0.935\pm0.008$ & $0.726\pm0.024$ & $154.3\pm0.6$ & 1.150 \\
1(c) & 3.6  & 12--23 & $p$ & $1.031\pm0.015$ & $0.180\pm0.010$ & $0.893\pm0.006$ & $0.802\pm0.030$ & $116.2\pm0.8$ & 2.838 \\
1(d) & 3.6  & 23--39 & $p$ & $1.033\pm0.013$ & $0.213\pm0.008$ & $0.818\pm0.012$ & $0.812\pm0.030$ & $80.1\pm0.5$  & 3.562 \\
1(e) & 3.6  & 39--81 & $p$ & $0.923\pm0.024$ & $0.232\pm0.012$ & $0.685\pm0.015$ & $0.753\pm0.054$ & $19.5\pm0.1$  & 1.422 \\
\hline
2(a) & 4.1  & 0--5   & $p$ & $1.043\pm0.018$ & $0.300\pm0.011$ & $0.895\pm0.012$ & $0.752\pm0.066$ & $190.1\pm1.8$ & 0.561 \\
2(b) & 4.1  & 5--12  & $p$ & $1.148\pm0.023$ & $0.378\pm0.015$ & $0.854\pm0.018$ & $0.858\pm0.084$ & $170.3\pm1.2$ & 0.573 \\
2(c) & 4.1  & 12--23 & $p$ & $1.025\pm0.020$ & $0.352\pm0.012$ & $0.765\pm0.010$ & $0.950\pm0.090$ & $124.2\pm1.0$ & 0.488 \\
2(d) & 4.1  & 23--39 & $p$ & $1.038\pm0.022$ & $0.323\pm0.013$ & $0.711\pm0.008$ & $1.026\pm0.024$ & $86.2\pm0.8$  & 0.812 \\
2(e) & 4.1  & 39--81 & $p$ & $1.024\pm0.024$ & $0.333\pm0.013$ & $0.565\pm0.008$ & $1.110\pm0.066$ & $25.5\pm0.1$  & 0.146 \\
\hline
3(a) & 4.7  & 0--5   & $p$ & $1.002\pm0.023$ & $0.338\pm0.017$ & $0.878\pm0.010$ & $0.966\pm0.066$ & $182.4\pm1.2$ & 0.820 \\
3(b) & 4.7  & 5--12  & $p$ & $1.042\pm0.018$ & $0.362\pm0.013$ & $0.865\pm0.007$ & $1.088\pm0.084$ & $165.2\pm1.0$ & 1.072 \\
3(c) & 4.7  & 12--23 & $p$ & $1.024\pm0.021$ & $0.387\pm0.010$ & $0.758\pm0.012$ & $1.242\pm0.060$ & $135.2\pm1.0$ & 2.049 \\
3(d) & 4.7  & 23--39 & $p$ & $1.006\pm0.016$ & $0.353\pm0.007$ & $0.682\pm0.008$ & $1.198\pm0.066$ & $87.1\pm0.7$  & 0.649 \\
3(e) & 4.7  & 39--77 & $p$ & $1.005\pm0.018$ & $0.387\pm0.013$ & $0.462\pm0.008$ & $1.240\pm0.054$ & $35.0\pm0.5$  & 1.441 \\
\hline
4(a) & 2.4  & 0--5   & $p$ & $0.583\pm0.008$ & $0.079\pm0.006$ & $0.989\pm0.006$ & $0.295\pm0.042$ & $118.3\pm1.3$ & 1.577 \\
4(b) & 3.1  & 0--5   & $p$ & $0.775\pm0.015$ & $0.123\pm0.010$ & $0.993\pm0.008$ & $0.612\pm0.036$ & $141.7\pm1.0$ & 1.381 \\
4(c) & 3.6  & 0--5   & $p$ & $0.966\pm0.018$ & $0.133\pm0.012$ & $0.992\pm0.007$ & $0.798\pm0.033$ & $152.5\pm0.8$ & 2.258 \\
4(d) & 4.1  & 0--5   & $p$ & $0.978\pm0.016$ & $0.130\pm0.014$ & $0.990\pm0.008$ & $0.809\pm0.033$ & $152.2\pm1.1$ & 0.570 \\
\hline
5(a) & 5    & 0--5   & $p-\bar p$ & $0.828\pm0.027$ & $0.288\pm0.015$ & $0.850\pm0.016$ & $0.903\pm0.060$ & $154.5\pm1.2$ & 1.088 \\
5(b) & 17.3 & 0--5   & $p-\bar p$ & $1.718\pm0.044$ & $0.685\pm0.034$ & $0.603\pm0.013$ & $1.393\pm0.069$ & $162.7\pm1.5$ & 0.431 \\
5(c) & 200  & 0--5   & $p-\bar p$ & $2.402\pm0.068$ & $0.727\pm0.041$ & $0.552\pm0.010$ & $2.580\pm0.078$ & $78.1\pm0.7$  & 0.171 \\
5(d) & 200  & 0--5   & $p$        & $2.255\pm0.072$ & $0.500\pm0.030$ & $0.866\pm0.014$ & $2.450\pm0.066$ & $171.5\pm1.1$ & 0.042 \\
\hline
6(a) & 8.8  & 0--5       & $p$ & $0.997\pm0.027$ & $0.596\pm0.016$ & $0.541\pm0.011$ & $1.166\pm0.078$ & $149.3\pm0.9$ & 0.215 \\
6(b) & 8.8  & 5--12.5    & $p$ & $1.125\pm0.030$ & $0.736\pm0.012$ & $0.350\pm0.013$ & $1.126\pm0.078$ & $130.5\pm1.0$ & 2.536 \\
6(c) & 8.8  & 12.5--23.5 & $p$ & $1.129\pm0.029$ & $0.722\pm0.022$ & $0.457\pm0.010$ & $1.533\pm0.069$ & $111.2\pm0.8$ & 0.413 \\
6(d) & 8.8  & 23.5--33.5 & $p$ & $1.070\pm0.030$ & $0.655\pm0.025$ & $0.360\pm0.015$ & $1.682\pm0.072$ & $87.2\pm0.8$  & 1.384 \\
6(e) & 8.8  & 33.5--43.5 & $p$ & $1.068\pm0.028$ & $0.605\pm0.017$ & $0.340\pm0.012$ & $1.652\pm0.066$ & $62.1\pm0.7$  & 0.945 \\
\hline
7(a) & 16.8 & 0--5       & $p-\bar p$ & $1.796\pm0.056$ & $0.390\pm0.020$ & $0.762\pm0.013$ & $1.564\pm0.072$ & $165.8\pm1.2$ & 2.236 \\
7(b) & 16.8 & 5--12.5    & $p-\bar p$ & $1.853\pm0.050$ & $0.524\pm0.024$ & $0.708\pm0.018$ & $1.722\pm0.066$ & $143.2\pm1.2$ & 1.136 \\
7(c) & 16.8 & 12.5--23.5 & $p-\bar p$ & $1.778\pm0.053$ & $0.483\pm0.017$ & $0.717\pm0.010$ & $1.506\pm0.078$ & $91.1\pm1.0$  & 0.442 \\
7(d) & 16.8 & 23.5--33.5 & $p-\bar p$ & $1.743\pm0.047$ & $0.523\pm0.017$ & $0.592\pm0.015$ & $1.744\pm0.078$ & $70.3\pm0.3$  & 2.014 \\
7(e) & 16.8 & 33.5--43.5 & $p-\bar p$ & $1.730\pm0.050$ & $0.562\pm0.018$ & $0.562\pm0.012$ & $1.766\pm0.072$ & $46.2\pm0.4$  & 0.094 \\
\hline
\end{tabular}%
\end{center}
}} }

\vskip0.5cm

Figures 4 and 5 are the same as Figure 1, but they show the
results of protons [Figures 4 and 5(d)] and net-protons [Figures
5(a)--5(c)] emitted in 0--5\% Au-Au collisions over a
$\sqrt{s_{NN}}$ range from 2.4 to 200 GeV, where the specific
energies are marked in the panels. The closed circles represent
the experimental data of the E895 [18], E802/E877/E917, NA49, and
BRAHMS Collaborations [7, 19], and the open circles are reflected
at the mid-rapidity. The curves are our results fitted by the
three-source distribution. The values of free parameters
$\sigma_C$, $\sigma_F$, $k_C$, and $\Delta y$, as well as the
normalization constant $N_0$ are listed in Table 1 with the values
of $\chi^2$/dof. One can see that the model describes
approximately the considered experimental data in most cases.

Figures 6 and 7 are similar to Figure 1, but they show the results
of protons and net-protons emitted in Pb-Pb collisions at
$\sqrt{s_{NN}}=8.8$ and 16.8 GeV, respectively. From panels (a) to
(d), the corresponding centrality intervals are 0--5\%, 5--12.5\%,
12.5--23.5\%, and 23.5--33.5\%, respectively. The closed circles
represent the experimental data of the NA49 Collaboration [20] and
the open circles are reflected at the mid-rapidity. The curves are
our results fitted by the three-source distribution. The values of
$\sigma_C$, $\sigma_F$, $k_C$, $\Delta y$, $N_0$, and $\chi^2$/dof
are listed in Table 1. One can see that the model describes
approximately the considered experimental data in most cases.

\begin{figure}
\hskip-1.0cm \begin{center}
\includegraphics[width=15.0cm]{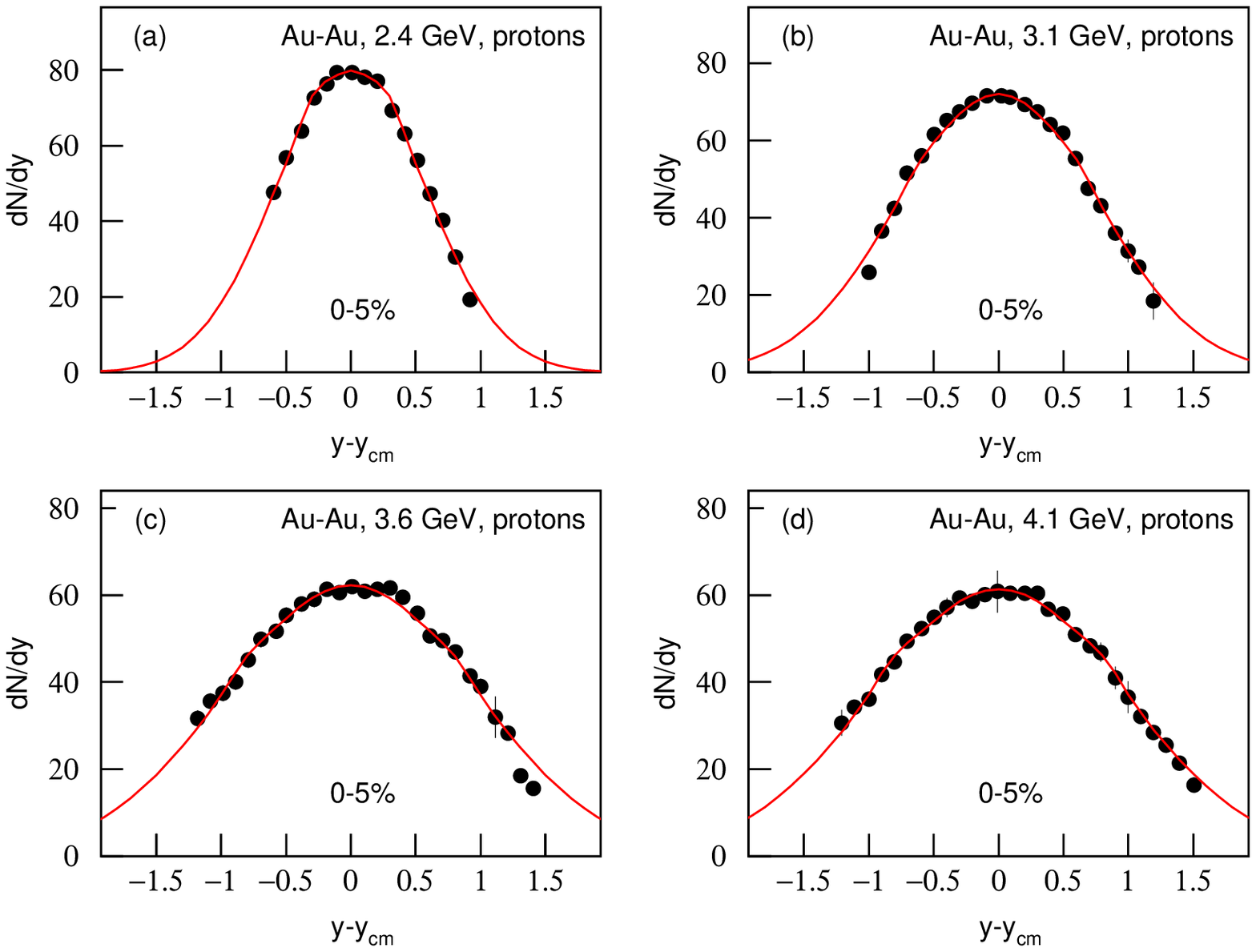}
\end{center}
\vskip.0cm\small Figure 4. Rapidity distributions of protons
emitted in 0--5\% Au-Au collisions at $\sqrt{s_{NN}}$ being (a)
2.4, (b) 3.1, (c) 3.6, and (d) 4.1 GeV, respectively. The closed
circles represent the experimental data of the E895 Collaboration
[18]. The curves are our results fitted by the three-source
distribution.
\end{figure}

We notice that there are differences between the rapidity
distributions of protons and net-protons. For instance, the data
in Figures 5(c) and 5(d) are obviously different from each other.
If the contributions of leading protons are excluded, the rapidity
distribution shapes of protons and antiprotons are similar [21,
22], and they have the same rules as a function of the collision
energy and centrality. Including the contributions of leading
protons in the rapidity distribution of protons, two peaks appear
in the forward and backward rapidity regions respectively, which
is obviously different from those of antiprotons. As a result of
protons minus antiprotons, the rapidity distribution of
net-protons shows large differences from that of protons in all
rapidity regions, particularly in very central and middle
forward/backward rapidity regions.

As for the collisions which have the same collision energy and
centrality, we can explain the differences by the statistical
fluctuations and distribution ranges. In particular, the
differences between Figures 4(c) and 1(a), as well as Figures 4(d)
and 2(a), could be caused by the statistical fluctuations in
different experiments measured by different collaborations in
different rapidity ranges [17, 18]. In our opinion, to give a good
enough description, one should use a wide enough rapidity
distribution. The rapidity range of the E895 data presented in
Figure 4 is wider than that of the E917 data presented in Figures
1--3. We are inclined to think that the description for Figure 4
is more accurate. In fact, we expect to describe the data in the
whole rapidity range. Although emulsion experiment can satisfy
this expectation, it is difficult to identify different particles
and to give high statistics.

To study the excitation functions of free parameters, the left
panel of Figure 8 shows the dependences of (a) $\sigma_C$, (c)
$\sigma_F$, (e) $k_C$, and (g) $\Delta y$ on $\sqrt{s_{NN}}$,
respectively. The closed and open symbols represent parameter
values from protons and net-protons respectively, and different
symbols are for different centrality intervals shown in the panel.
All parameter values are extracted from Figures 1--7 and listed in
Table 1. It should be noticed that the error bars look very small
due to all the four parameters being very sensitive quantities for
rapidity distribution. A very small increase or decrease in the
parameter values can cause a large dispersion from the
expectation. To obtain a reasonable fitting result, we have to
limit the parameters in very small ranges. The line segments in
Figures 8(a) and 8(c) for central collisions are used to guide the
eyes. Only the results for central collisions are presented in
Figures 8(e) and 8(g) for the purpose of clarity and concision,
where the solid and dashed lines in Figure 8(g) are our results
fitted by the linear functions
\begin{equation}
\Delta y=(0.080\pm0.021)+(0.470\pm0.067)\ln(\sqrt{s_{NN}})
\end{equation}
with $\chi^2$/dof=6.862 and
\begin{equation}
\Delta y=(0.095\pm0.027)+(0.483\pm0.060)\ln(\sqrt{s_{NN}})
\end{equation}
with $\chi^2$/dof=2.324 respectively.

In the case of considering protons and net-protons together, from
the left panel of Figure 8 one can see that $\sigma_C$ increases
generally with increase of $\sqrt{s_{NN}}$. At the same time,
$\sigma_F$ increases initially with increase of $\sqrt{s_{NN}}$
and then saturates at above 8.8 GeV and fluctuates around its
value at 17 GeV. As the parameter described the fraction of the
central rapidity region, $k_C$ seems to show a confused energy
dependence. In fact, $k_C$ is related to other parameters which
depend on energy. A careful observation shows a tendency that,
with increasing $\sqrt{s_{NN}}$, $k_C$ decreases quickly from
$0.98$ at 2.1 GeV to $0.55$ at 8.8 GeV, and saturates to about
$0.70$ at above 17 GeV. This renders that $k_C$ has a minimum at
$\sqrt{s_{NN}}=8.8$ GeV. In our considered energy range, $\Delta
y$ increases linearly with increase of $\ln(\sqrt{s_{NN}})$. Our
observation shows the protons and net-protons having consistent
results in most cases.

The characteristics of the excitation functions have some possible
reasons. The trend of general increase in $\sigma_C$ renders a
continuous expansion of the central rapidity region in the
considered energy range. The saturation of $\sigma_F$ at
$\sqrt{s_{NN}}>8.8$ GeV renders the scaling law and nuclear
limiting fragmentation in the forward/backward rapidity regions.
The saturation of $k_C$ at $\sqrt{s_{NN}}>17$ GeV renders that the
number of non-leading protons which stay in the central rapidity
region are independent of energy at higher AGS and RHIC energies,
and the scaling law and nuclear limiting fragmentation appear
again. The linear increase in $\Delta y$ reflects the power of
nuclear penetration which also increases with increase of
$\ln(\sqrt{s_{NN}})$.

To study the relations of free parameters and the collision
centrality $C$, the right panel of Figure 8 shows the dependences
of (b) $\sigma_C$, (d) $\sigma_F$, (f) $k_C$, and (h) $\Delta y$
on $C$, respectively. The closed and open symbols represent
parameter values for protons and net-protons respectively, and
different symbols are for different energies shown in the panel,
which are extracted from Figures 1--7 and listed in Table 1. As
mentioned in the above paragraph on the left panel of Figure 8,
because all the four parameters are very sensitive quantities for
rapidity distribution, they show very small error bars. The lines
in Figures 8(b), 8(d), and 8(f) are our results fitted by a linear
function, though some of them do not sure obey the linear
relation. The corresponding intercepts and slopes are listed in
Table 2 with $\chi^2$/dof, where we see a few large $\chi^2$/dof
which imply a non-linear relation. The line segments in Figure
8(h) are used to guide the eyes.

\begin{figure}
\hskip-1.0cm \begin{center}
\includegraphics[width=15.0cm]{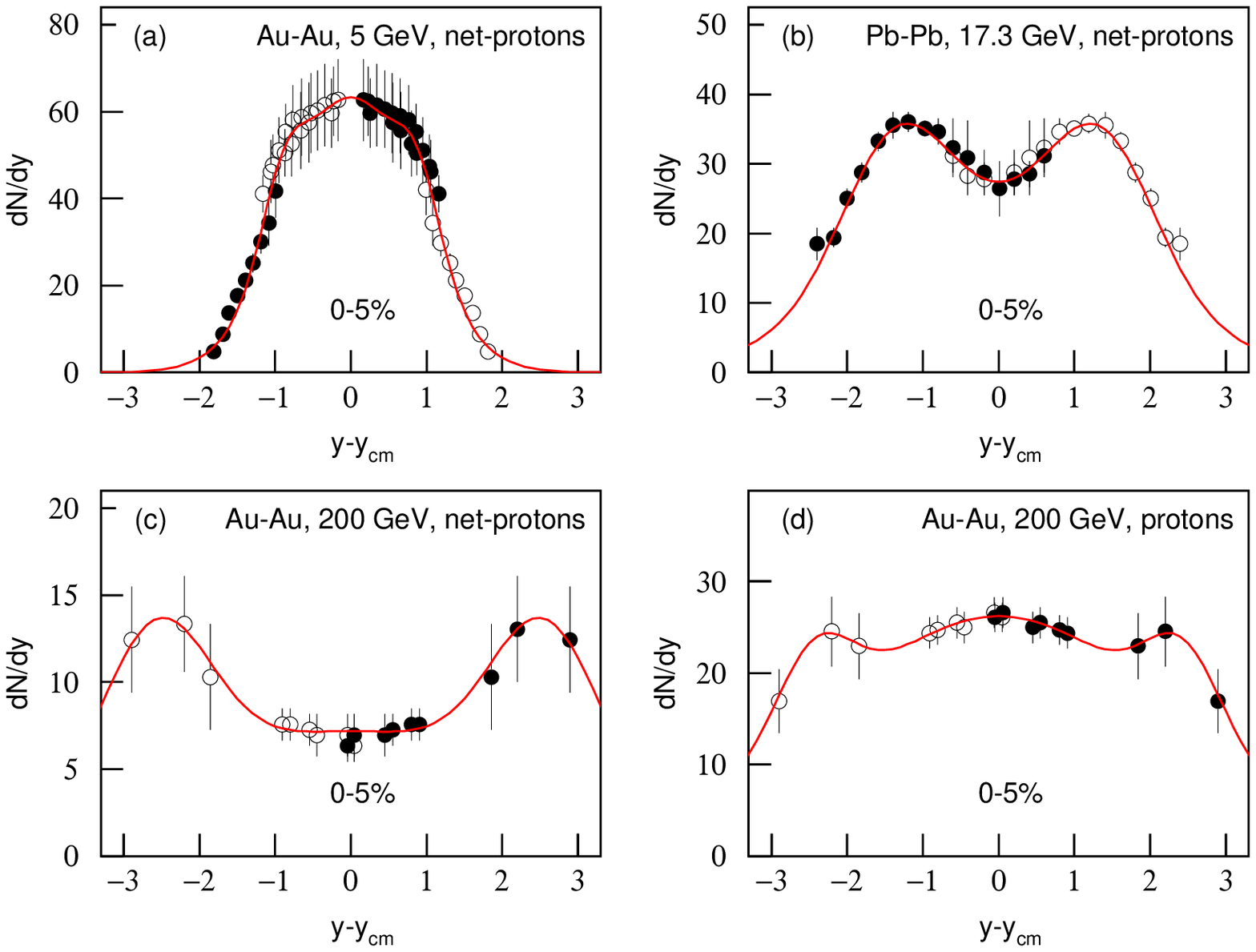}
\end{center}
\vskip.0cm\small Figure 5. The same as Figure 4, but showing the
results of (a)--(c) net-protons and (d) protons emitted in 0--5\%
Au-Au collisions at $\sqrt{s_{NN}}$ being (a) 5, (b) 17.3, and
(c)(d) 200 GeV, respectively. The closed circles represent the
experimental data of the E802/E877/E917, NA49, and BRAHMS
Collaborations [7, 19] and the open circles are reflected at the
mid-rapidity.
\end{figure}

\begin{figure}
\hskip-1.0cm \begin{center}
\includegraphics[width=15.0cm]{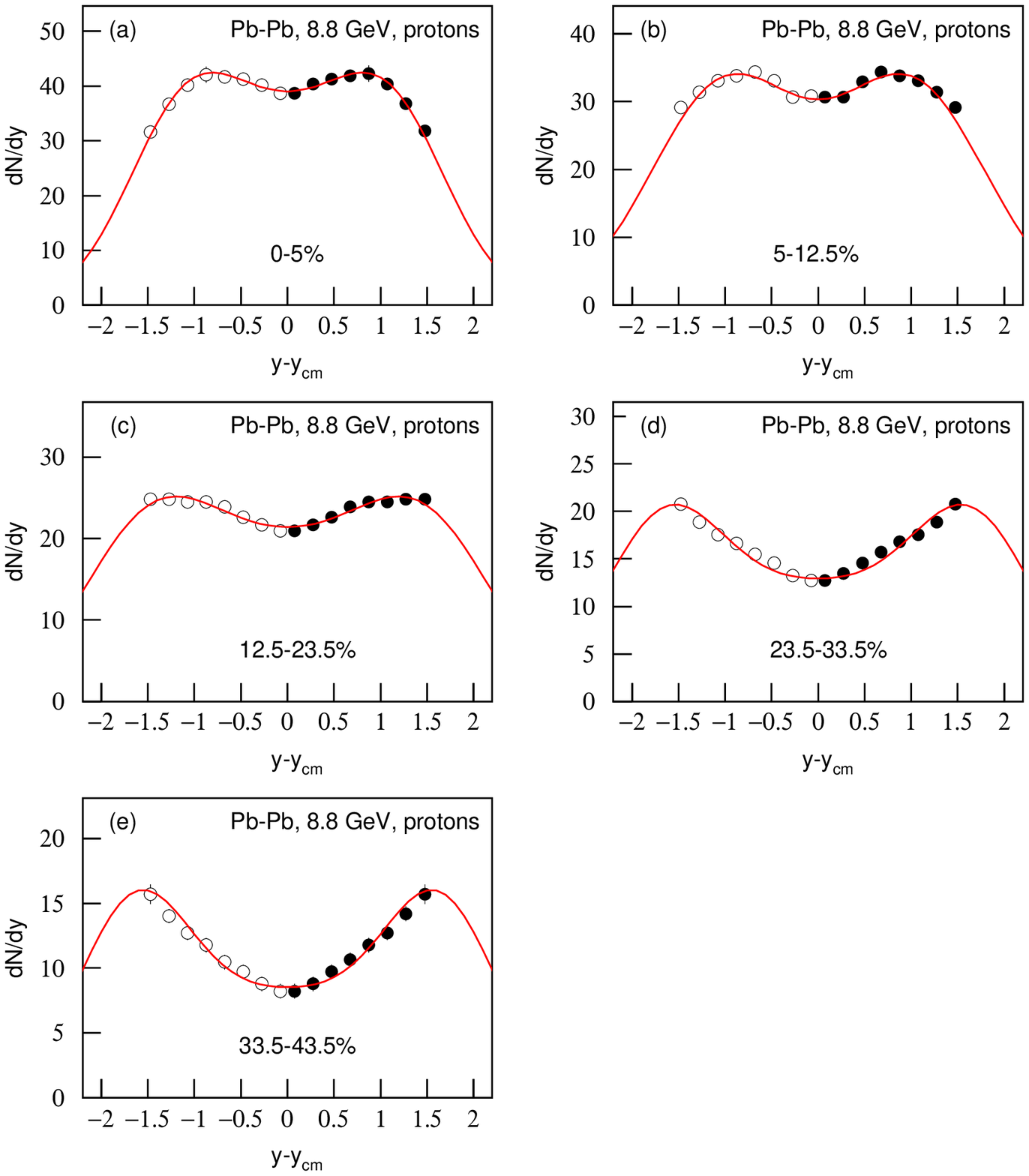}
\end{center}
\vskip.0cm\small Figure 6. Rapidity distributions of protons
emitted in Pb-Pb collisions at $\sqrt{s_{NN}}=8.8$ GeV, where
panels (a)--(d) correspond to centrality intervals 0--5\%,
5--12.5\%, 12.5--23.5\%, and 23.5--33.5\%, respectively. The
closed circles represent the experimental data of the NA49
Collaboration [20] and the open circles are reflected at the
mid-rapidity. The curves are our results fitted by the
three-source distribution.
\end{figure}

\begin{figure}
\hskip-1.0cm \begin{center}
\includegraphics[width=15.0cm]{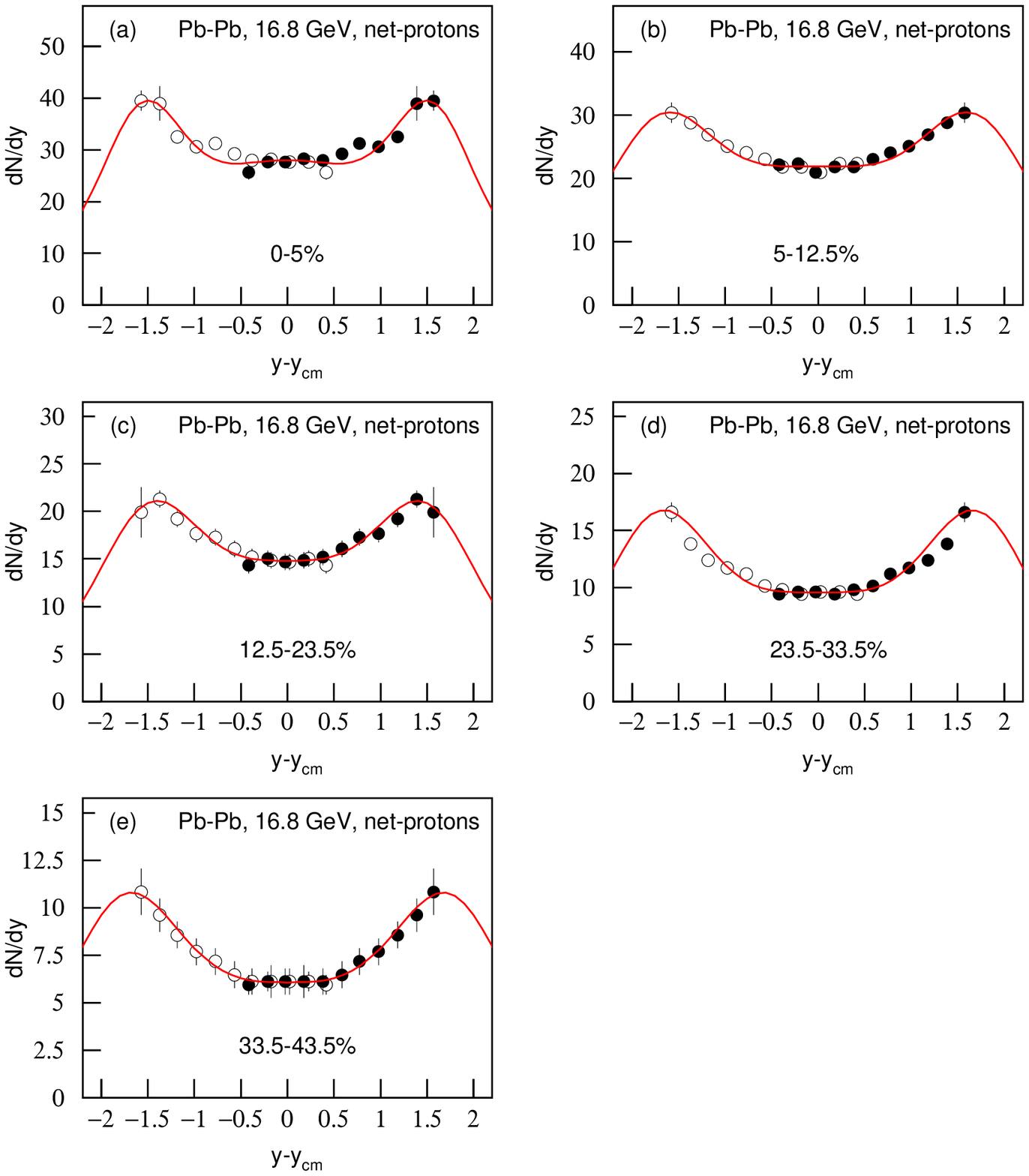}
\end{center}
\vskip.0cm\small Figure 7. The same as Figure 6, but showing the
results of net-protons emitted in Pb-Pb collisions at
$\sqrt{s_{NN}}=16.8$ GeV.
\end{figure}

\begin{figure}
\hskip-1.0cm \begin{center}
\includegraphics[width=15.0cm]{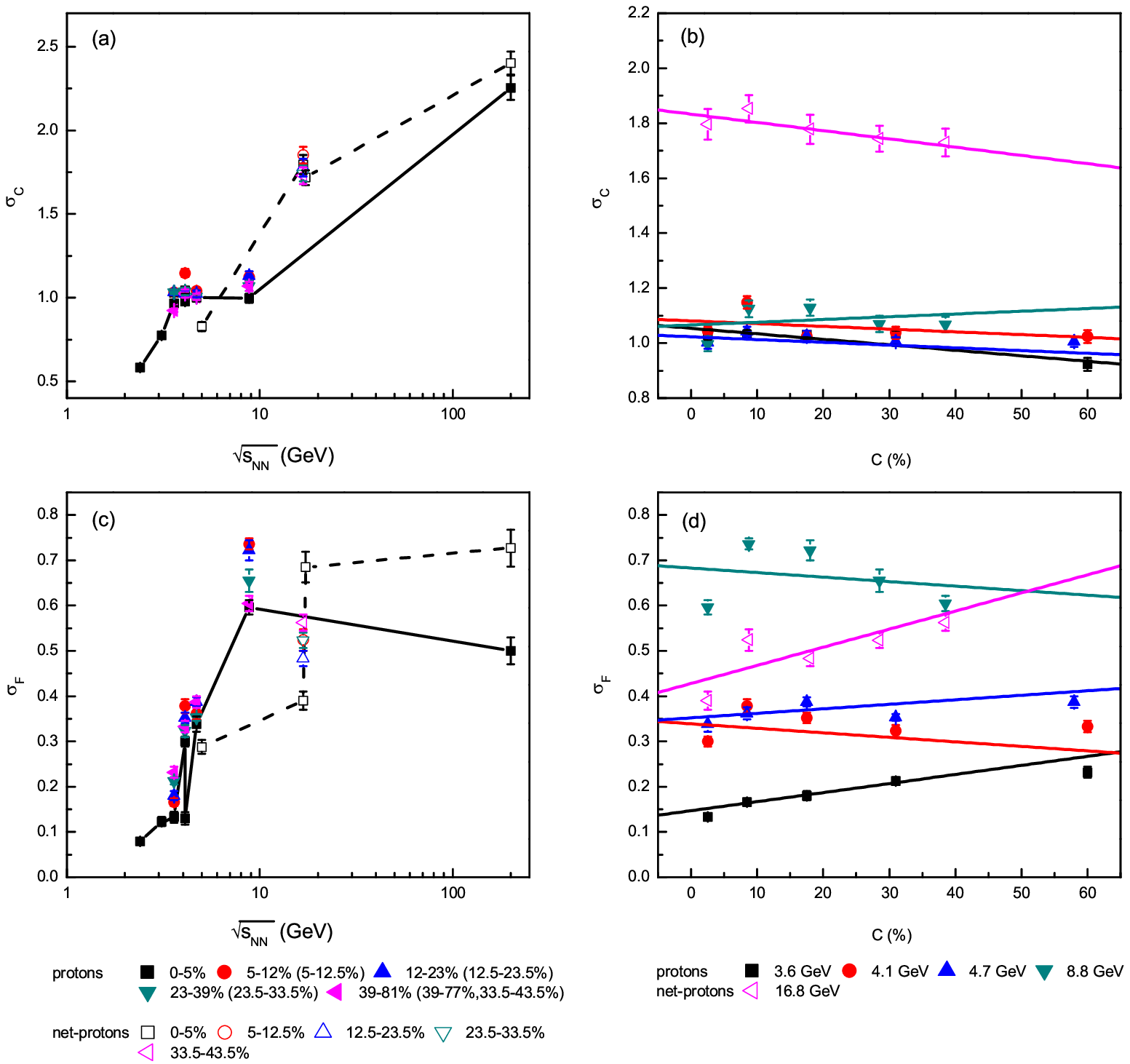}
\end{center}
\vskip.0cm\small Figure 8. {\it Left:} Excitation functions of (a)
$\sigma_C$, (c) $\sigma_F$, (e) $k_C$, and (g) $\Delta y$.
Different symbols represent parameter values for different
centrality intervals, where the closed and open symbols are from
protons and net-protons respectively. The line segments in Figures
8(a) and 8(c) for central collisions are used to guide the eyes.
Only the results for central collisions are presented in Figures
8(e) and 8(g) for the purpose of clarity and concision. The lines
in Figure 8(g) are our results fitted by linear functions. {\it
Right:} Dependences of (b) $\sigma_C$, (d) $\sigma_F$, (f) $k_C$,
and (h) $\Delta y$ on the centrality $C$. Different symbols
represent parameter values for different energies, where the
closed and open symbols are from protons and net-protons
respectively. The lines in Figures 8(b), 8(d), and 8(f) are our
results fitted by linear functions. The line segments in Figure
8(h) are used to guide the eyes.
\end{figure}

\begin{figure}
\hskip-1.0cm \begin{center}
\includegraphics[width=15.0cm]{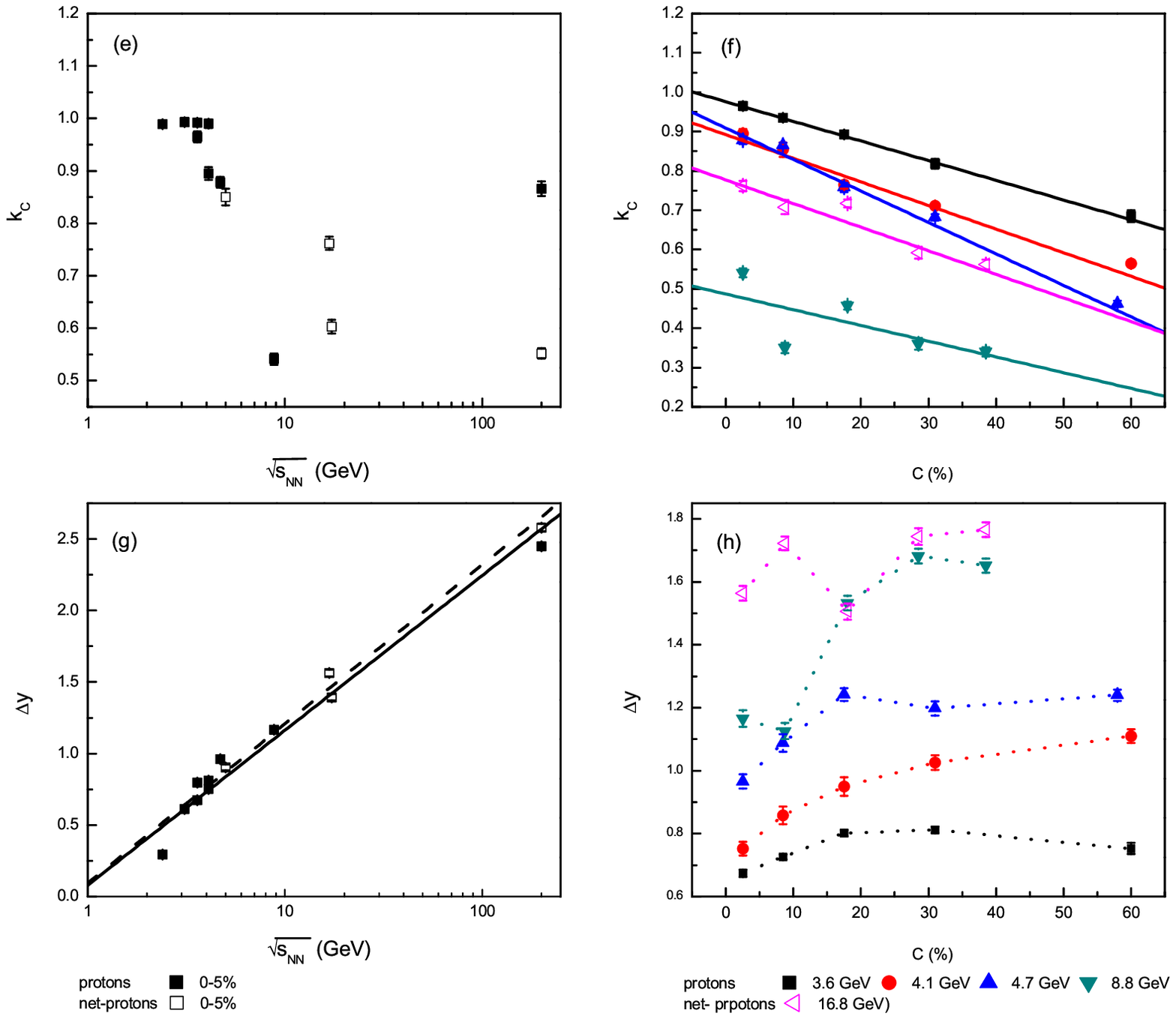}
\end{center}
\vskip.0cm\small Figure 8. Continued.
\end{figure}

One can see from the right panel of Figure 8 that, for the
considered five energies, $\sigma_C$ and $\sigma_F$ have a slight
change or do not change with decrease of the centrality (or
increase of the centrality percentage). With decrease of the
centrality, $k_C$ decreases conformably for the five energies, and
$\Delta y$ increases for the energies at or near the minimum and
does not change obviously for the energies apart from the minimum.

\vskip1.0cm

{\small {Table 2. Values of intercept, slope, and $\chi^2$/dof
corresponding to the lines in Figures 8(b), 8(d), and 8(f).
{%
\begin{center}
\begin{tabular}{cccccc}
\hline Figure & Relation & $\sqrt{s_{NN}}$ (GeV) & Intercept & Slope & $\chi^2$/dof \\
\hline
8(b) & $\sigma_C-C$ & 3.6  & $1.054\pm0.016$ & $-0.002\pm0.001$ & 3.466 \\
     &              & 4.1  & $1.081\pm0.032$ & $-0.001\pm0.001$ & 6.190 \\
     &              & 4.7  & $1.023\pm0.011$ & $-0.001\pm0.001$ & 0.842 \\
     &              & 8.8  & $1.066\pm0.042$ & $0.001\pm0.001$  & 4.674 \\
     &              & 16.8 & $1.833\pm0.022$ & $-0.003\pm0.001$ & 0.380 \\
8(d) & $\sigma_F-C$ & 3.6  & $0.147\pm0.010$ & $0.002\pm0.001$  & 4.127 \\
     &              & 4.1  & $0.339\pm0.020$ & $-0.001\pm0.001$ & 7.344 \\
     &              & 4.7  & $0.352\pm0.012$ & $0.001\pm0.001$  & 4.353 \\
     &              & 8.8  & $0.683\pm0.050$ & $-0.001\pm0.002$ & 22.138 \\
     &              & 16.8 & $0.428\pm0.032$ & $0.004\pm0.001$  & 4.415 \\
8(f) & $k_C-C$      & 3.6  & $0.976\pm0.003$ & $-0.005\pm0.001$ & 0.167 \\
     &              & 4.1  & $0.892\pm0.013$ & $-0.006\pm0.001$ & 4.322 \\
     &              & 4.7  & $0.909\pm0.011$ & $-0.008\pm0.001$ & 4.794 \\
     &              & 8.8  & $0.487\pm0.051$ & $-0.004\pm0.002$ & 38.062 \\
     &              & 16.8 & $0.777\pm0.021$ & $-0.006\pm0.001$ & 7.063 \\
\hline
\end{tabular}%
\end{center}
}} }

\vskip0.5cm

The characteristics of the dependences of parameters on the
centrality imply some possible reasons. The near independences of
$\sigma_C$ ($\sigma_F$) and partly $\Delta y$ from $C$ render that
the rapidity shift of (net-)protons are mainly determined by the
collision energy. At or near the minimums, less $\Delta y$ in
central collisions renders larger viscosity due to more
participant nucleons in the collisions. The decrease trend of
$k_C$ on $C$ renders that multi-scattering results in more
non-leading protons in central collisions than in peripheral
collisions. The leading protons and their number determine both
the rapidity shift and fraction in given central and forward
(backward) rapidity regions.

We would like to point out that the parameters $\sigma_C$
($\sigma_F$) and $\Delta y$ are generally independent of, and the
parameter $k_C$ is dependent on, the types of protons and
net-protons. In particular, in Figure 8(e), there are a few points
which are taken from the rapidity distributions of net-protons at
$\sqrt{s_{NN}}=5$, 16.8, 17.3, and 200 GeV, which give smaller
$k_C$ than that from protons and do not affect our main conclusion
on the dependence of $k_C$ on $\sqrt{s_{NN}}$. If all of the
points at $\sqrt{s_{NN}}=5$, 16.8, 17.3, and 200 GeV are taken
from protons, we expect a higher saturation value ($>0.70$) for
$k_C$ at $\sqrt{s_{NN}}>17$ GeV. In Figure 8(f), the points at
$\sqrt{s_{NN}}=16.8$ GeV are taken from net-protons, which show a
similar law as those taken from protons at lower energies.

The present work confirms one of our recent works [16] in which
the method of squared speed-of-sound analysis is used according to
the widths of rapidity distributions of negatively charged pions,
and the minimum is observed at $\sqrt{s_{NN}}=8.8$ GeV. The
present work also confirms another recent work of ours [23] in
which the method of string tension analysis is used due to the
Schwinger mechanism description of transverse momentum
distribution of identified particles, and the minimum is observed
at $\sqrt{s_{NN}}=7.7$ GeV.

The present work is consistent with other works. For example,
refs. [24, 25] indicated the minimum locating in $\sqrt{s_{NN}}$
range from 4 to 9 GeV, in which the Landau hydrodynamic model and
the ultra-relativistic quantum molecular dynamics hybrid approach
are used. A wiggle in the excitation function of a specific
reduced curvature of the net-proton rapidity distribution at
midrapidity is expected in the $\sqrt{s_{NN}}$ range from 4 to 8
GeV [4--6]. Other works which study the excitation functions of
mean transverse mass minus rest mass [26], chemical freeze-out
temperature [26, 27], yield ratios of positive kaons to pions
[26--28], and width ratios of experimental negative pion rapidity
distribution to Landau hydrodynamic model prediction [27] show a
knee point around $\sqrt{s_{NN}}=7$--8 GeV.

It is hard to connect the minimums of the model parameters with
the softest points of the EoS, because the three-source
distribution used in the present work does not consider the system
evolution and the EoS. Various minimums extracted from different
excitation functions may be different. To search for the minimal
minimum, we have to extract the minimums of the excitation
functions of the parameters in many distributions and correlations
which include, but are not limited to, the rapidity distribution,
transverse momentum spectrum, multiplicity distribution,
transverse energy spectrum, fragmentation function distribution,
dependences of anisotropic flow on rapidity and transverse
momentum, and anisotropic flow distribution. This is a huge
project which is beyond the scope of the present work.

It should be noted that, in our previous works [14--16], the width
of rapidity distribution of newly produced particles is used to
extract the squared speed-of-sound parameter based on the Landau
hydrodynamic model [29--31]. Protons and net-protons are not newly
produced particles. They are confined by the baryon conservation.
Strictly speaking, the (net-)protons are beyond the scope of the
Landau hydrodynamic model. In fact, this model does not consider
the continuity equation for baryon charge conservation. Therefore,
the width of rapidity distribution of (net-)protons discussed in
the present work should not be used to extract the squared
speed-of-sound parameter.
\\

{\section{Conclusions}}

We summarize here our main observations and conclusions.

a) The rapidity distributions of protons and net-protons emitted
in Au-Au and Pb-Pb collisions over a $\sqrt{s_{NN}}$ range from
2.4 to 200 GeV are analyzed by using the three-source
distribution. The model results can fit the experimental data of a
few collaborations who worked at the AGS, SPS, and RHIC,
respectively. Some interesting trends of parameters on energy are
obtained.

b) The rapidity distribution width $\sigma_C$ for the central
rapidity region increases generally with increase of
$\sqrt{s_{NN}}$. The trend of general increase in $\sigma_C$
renders a continuous expansion of the central rapidity region in
the considered energy range. The rapidity distribution width
$\sigma_F$ for the forward/backward rapidity regions increases
initially with increase of $\sqrt{s_{NN}}$ and then saturates at
above 8.8 GeV and fluctuates around its value at 17 GeV. The
saturation of $\sigma_F$ renders the nuclear limiting
fragmentation in the forward/backward rapidity regions.

c) With increasing $\sqrt{s_{NN}}$, $k_C$ decreases quickly from
$0.98$ at 2.1 GeV to $0.55$ at 8.8 GeV, and saturates to about
$0.70$ at above 17 GeV. This renders that $k_C$ has a minimum at
$\sqrt{s_{NN}}=8.8$ GeV. The saturation of $k_C$ renders that the
number of non-leading protons which stay in the central rapidity
region is independent of energy at $\sqrt{s_{NN}}>17$ GeV. In the
considered energy range, $\Delta y$ increases linearly with
increase of $\ln(\sqrt{s_{NN}})$, which reflects the power of
nuclear penetration which also increases with increase of
$\ln(\sqrt{s_{NN}})$.

d) The minimum in the excitation function of $k_C$ obtained in the
present work confirms one of our recent works [16] in which the
minimum is observed at $\sqrt{s_{NN}}=8.8$ GeV by using the method
of squared speed-of-sound analysis on the widths of rapidity
distributions of negatively charged pions. The present work also
confirms another recent work ours [23] in which the minimum is
observed at $\sqrt{s_{NN}}=7.7$ GeV by using the method of string
tension analysis in the Schwinger mechanism on the transverse
momentum distribution of identified particles.

e) Mainly for the considered five energies, $\sigma_C$ and
$\sigma_F$ have a slight change or do not change with decrease of
the centrality (or increase of the centrality percentage). With
decrease of the centrality, $\Delta y$ increases for the energies
at or near the minimums, and does not change obviously for the
energies apart from the minimums. The near independences of
$\sigma_C$ ($\sigma_F$) and partly $\Delta y$ from centrality
render that the rapidity shift of (net-)protons are mainly
determined by the collision energy. At or near the minimums, less
$\Delta y$ in central collisions renders larger viscosity due to
more participant nucleons in the collisions.

f) With decrease of the centrality, $k_C$ decreases conformably
for the five energies. The decrease trend of $k_C$ on centrality
renders that multi-scattering results in more non-leading protons
in central collisions than in peripheral collisions. The leading
protons and their number determine both the rapidity shift and
fraction in given central and forward (backward) rapidity regions.
\\

{\bf Conflict of Interests}

The authors declare that there is no conflict of interests
regarding the publication of this paper.
\\

{\bf Acknowledgments}

This work was supported by the National Natural Science Foundation
of China under Grant No. 11575103 and the US DOE under contract
DE-FG02-87ER40331.A008.


\vskip1.0cm

\begin{thebibliography}{99}
\setlength{\itemsep}{-1pt}
\bibitem{1}
EMU-01 Collaboration (M.I. Adamovich et al.), Phys. Lett. B {\bf
201}, 397 (1988).
\bibitem{2}
K.K. Olimov, Q. Ali, M.Q. Haseeb, A. Arif, S.L. Lutpullaev, B.S.
Yuldashev, Int. J. Mod. Phys. E {\bf 24}, 1550049 (2015).
\bibitem{3}
Z.J. Jiang, J. Wang, Y. Huang, Int. J. Mod. Phys. E {\bf 25},
1650025 (2016).
\bibitem{4}
Y.B. Ivanov and D. Blaschke, Phys. Rev. C {\bf 92}, 024916 (2015).
\bibitem{5}
Y.B. Ivanov, Phys. Lett. B {\bf 721}, 123 (2013).
\bibitem{6}
Y.B. Ivanov,  Phys. Rew. C {\bf 87}, 064904 (2013).
\bibitem{7}
BRAHMS Collaboration (I.G. Bearden et al.), Phys. Rev. Lett. {\bf
93}, 102301 (2004).
\bibitem{8}
R.A. Lacey, Phys. Rev. Lett. {\bf 114}, 142301 (2015).
\bibitem{9}
NA49 Collaboration (S.V. Afanasiev et al.), Phys. Rev. C {\bf 66},
054902 (2002).
\bibitem{10}
NA49 Collaboration (C. Alt et al.), Phys. Rev. C {\bf 77}, 024903
(2008).
\bibitem{11}
G. Wolschin, Eur. Phys. J. A {\bf 5}, 85 (1999).
\bibitem{12}
G. Wolschin, J. Phys. G {\bf 40}, 045104 (2013).
\bibitem{13}
G. Wolschin, Prog. Part. Nucl. Phys. {\bf 59}, 374 (2007).
\bibitem{14}
L.-N. Gao, F.-H. Liu, Adv. High Energy Phys. {\bf 2015}, 184713
(2015).
\bibitem{15}
L.-N. Gao, F.-H. Liu, Adv. High Energy Phys. {\bf 2015}, 641906
(2015).
\bibitem{16}
F.-H. Liu, L.-N. Gao, R.A. Lacey, Adv. High Energy Phys. {\bf
2016}, 9467194 (2016).
\bibitem{17}
E917 Collaboration (B.B. Back et al.), Phys. Rev. Lett. {\bf 86},
1970 (2001).
\bibitem{18}
E895 Collaboration (J.L. Klay et al.), Phys. Rev. Lett. {\bf 88},
102301 (2002).
\bibitem{19}
F. Videb{\ae}k, Nucl. Phys. A {\bf 830}, 43c (2009) (for the
BRAHMS collaboration).
\bibitem{20}
NA49 Collaboration (T. Anticic et al.), Phys. Rev. C {\bf 83},
014901 (2011).
\bibitem{21}
M. Murray, J. Phys. G {\bf 31}, S1137 (2005) (for the BRAHMS
Collabration).
\bibitem{22}
F.-H. Liu, Mod. Phys. Lett. A {\bf 23}, 337 (2008).
\bibitem{23}
L.-N. Gao, F.-H. Liu, R.A. Lacey, Eur. Phys. J. A {\bf 52}, 137
(2016).
\bibitem{24}
M. Bleicher, arXiv: hep-ph/0509314 (2005).
\bibitem{25}
J. Steinheimer, M. Bleicher, Eur. Phys. J. A {\bf 48}, 100 (2012).
\bibitem{26}
L. Kumar, J. Phys. G {\bf 38}, 124145 (2011) (for the STAR
Collaboration).
\bibitem{27}
A. Rustamov, Cent. Eur. J. Phys. {\bf 10}, 1267 (2012).
\bibitem{28}
D.T. Larsen, J. Phys. Conf. Ser. {\bf 668}, 012020 (2016).
\bibitem{29}
L.D. Landau, {\it Collected Papers of L. D. Landau}, D. Ter-Haarp
(Ed.), p. 569, Pergamon, Oxford, UK (1965).
\bibitem{30}
E.V. Shuryak, Yadernaya Fizika {\bf 16}, 395 (1972).
\bibitem{31}
P. Carruthers, Annals of the New York Academy of Sciences {\bf
229}, 91 (1974).

\end{thebibliography}
\end{document}